\documentclass[epj]{svjour}
\usepackage{graphics}
\usepackage{epsfig}
\begin{document}
\title{First CNGS events detected by LVD}
\author{
N.Yu.Agafonova \inst{1},
M.Aglietta \inst{2},
P.Antonioli \inst{3},
G.Bari \inst{3},
A.Bonardi\inst{2},
V.V.Boyarkin\inst{1},
G.Bruno\inst{2,4},
W.Fulgione\inst{2},
P.Galeotti\inst{2},
M.Garbini\inst{3,5},
P.L.Ghia\inst{2,4},
P.Giusti\inst{3},
E.Kemp\inst{6},
V.V.Kuznetsov\inst{1},
V.A.Kuznetsov\inst{1},
A.S.Malguin\inst{1},
H.Menghetti\inst{3},
R.Persiani\inst{3},
A.Pesci\inst{3},
I.A.Pless\inst{7},
A.Porta\inst{2},
V.G.Ryasny\inst{1},
O.G.Ryazhskaya\inst{1},
O.Saavedra\inst{2},
G.Sartorelli\inst{3},
M.Selvi\inst{3}\thanks{\emph{Corresponding author:} selvi@bo.infn.it},
C.Vigorito\inst{2},
L.Votano\inst{8},
V.F.Yakushev\inst{1},
G.T.Zatsepin\inst{1},
A.Zichichi\inst{3}.
%First author\inst{1} \and Second author\inst{2}% etc
}                     % Do not remove
\institute{
Institute for Nuclear Research, Russian Academy of Sciences, Moscow, Russia 
\and Institute of Physics of Interplanetary Space, INAF, Torino, University of Torino and INFN-Torino, Italy 
\and University of Bologna and INFN-Bologna, Italy
\and INFN-LNGS, Assergi, Italy
\and Museo Storico della Fisica, Centro Studi e Ricerche "E. Fermi", Rome, Italy
\and University of Campinas, Campinas, Brazil
\and Massachusetts Institute of Technology, Cambridge, USA
\and INFN-LNF, Frascati, Italy}
%Insert the first address here \and the second here}
%
\date{Received: date / Revised version: date}
\abstract{
The CERN Neutrino to Gran Sasso (CNGS) project aims to produce a high energy, wide band $\nu_{\mu}$ beam at CERN and send it toward the INFN Gran Sasso National Laboratory (LNGS), $732$ km away. Its main goal is the observation of the $\nu_{\tau}$ appearance, through neutrino flavour oscillation.
The beam started its operation in August 2006 for about 12 days: a total amount of $7.6~10^{17}$ protons were delivered to the target. The LVD detector, installed in hall A of the LNGS and mainly dedicated to the study of supernova neutrinos, was fully operating during the whole CNGS running time. A total number of $569$ events were detected in coincidence with the beam spill time. This is in good agreement with the expected number of events from Montecarlo simulations. 
 % end of PACS codes
} %end of abstract

\authorrunning{LVD collaboration}
\maketitle

\section{Introduction}                %% Introduction
The CERN Neutrinos to Gran Sasso (CNGS) project aims to produce a high energy, wide band $\nu_{\mu}$ beam at CERN and send it toward the INFN Gran Sasso National Laboratory (LNGS). Its main goal is the observation of the $\nu_{\tau}$ appearance, through neutrino flavour oscillation, by the Opera experiment.

The LVD detector, installed in the Hall A of the LNGS, is mainly dedicated to the observation of supernova neutrinos. As proven in \cite{LVDmonitor}, due to its large area and active mass, LVD can act as a very useful beam monitor, detecting the interaction of neutrinos inside the detector and the muons generated by the $\nu$ interaction in the rock upstream the detector.

The CNGS beam started its operation in August 2006, after three commissioning weeks. LVD was fully operative during the whole first run of the CNGS beam.

In this work we present the results of the events detected in coincidence with the beam spill time and show some comparisons with the Montecarlo simulation.

\section{The CNGS beam}               %% As many sections as you want
The informations about the CNGS beam characteristics are taken by the LHCLOG$\_$CNGS$\_$OPERA database (hereafter DB) \cite{db}. Two main quantities are relevant for each proton extraction: 
\begin{itemize}
\item the UTC time of the spill (in ns),
\item the number of extracted protons on target (p.o.t.)
\footnote{Due to some known problems (see \cite{DarioGS}) it happened that for some extraction there was only the UTC time or only the number of p.o.t. . In the following we will consider only those extraction where both the informations were present.}.
\end{itemize}

\begin{figure}[h!]
\resizebox{0.5\textwidth}{!}{%
  \includegraphics{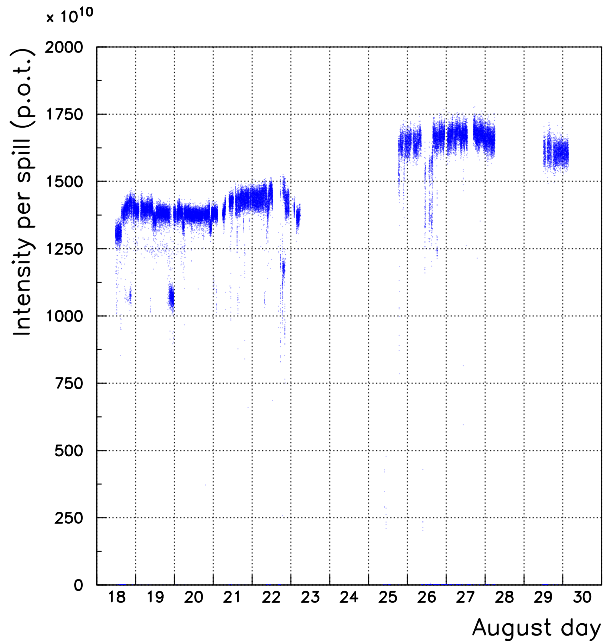}
}
%\mbox{\epsfig{file=intensity-per-spill.eps,height=10cm,width=10cm}}   %% EPS Figures
\caption{Beam intensity (in protons on target) per each spill of Run1.}
\label{fi:pot-time}
\end{figure}

The CNGS beam started its first run of operation on $18^{th}$ August, 2006 (first spill at 11:31:54.072 UTC) and finished on $30^{th}$ August (last spill at 03:30:04.872 UTC). In the following we will refer to this first run as Run1. The total number of protons delivered against the graphite target is $7.59~10^{17}$. 
%%%% COMMENT $7.82~10^{17}$ considering all the pot in the database even if there is no time associated. 
The beam intensity per each spill is shown in figure \ref{fi:pot-time}. 
It started with an average intensity of $1.38~10^{13}$ p.o.t. per spill until $23^{rd}$ August; then there was a predefined stop called ``Machine Development'' (MD). On $25^{th}$ August the run restarted with a slightly higher intensity of $1.64~10^{13}$ p.o.t. per spill, on average.

The time structure of the CNGS beam is characterized by two extractions, $10.5~\mu$s long, separated by $50$ ms; this pattern is repeated each CNGS cycle, whose duration can change: during Run1 there were two main repetition cycles: $16.8$ s and $22.8$ s.

\nopagebreak

\section{The LVD detector}               %% As many sections as you want
LVD is a large volume liquid scintillator detector dedicated to the study of core collapse supernova neutrinos. The active scintillator mass is about $1000$ t, while the iron and stainless steel support structure is $\sim 900$ t.
It has a modular structure, made of 840 identical scintillation counters. Each counter is viewed on the top by three photomultiplier tubes. The counters are grouped in three big modules, called ``towers'', with independent data acquisition. The energy calibration of the scintillation counters is done each month through the (known) energy released by cosmic muons. The highest detectable energy in each counter is between $200$ and $400$ MeV.

\begin{figure}[h!]
\resizebox{0.5\textwidth}{!}{%
  \includegraphics{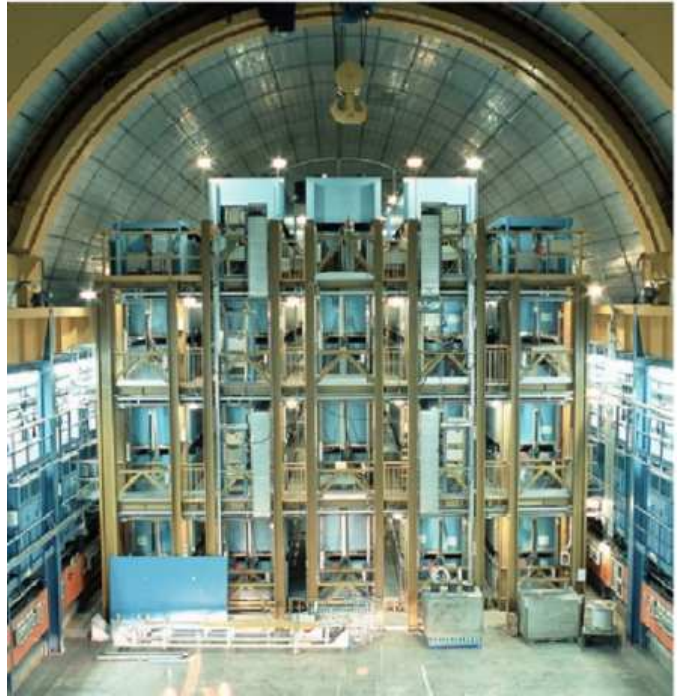}
}
%\mbox{\epsfig{file=TUNNEL2_zoom.eps,height=10cm,width=10cm}}   %% EPS Figures
\caption{Front view of the LVD detector.}
\label{fi:lvd}
\end{figure}

The front area of the whole detector (orthogonal to the CNGS beam) is $\sim 12 \times 10$ m$^2$, see figure \ref{fi:lvd}.
A detailed description of the detector and its performances is in \cite{lvd}.

During Run1 LVD was fully operative with $100\%$ of uptime (defined as the fraction of time where LVD is able to detect an event) and an average active mass of $950$ t.

\nopagebreak

\section{MC simulation of the expected events}
The CNGS events in LVD can be subdivided into two main categories:
\begin{itemize}
\item $\nu_{\mu}$ charged current (CC) interactions in the rock upstream the LNGS; they produce a muon that can reach LVD and be detected,
\item $\nu_{\mu}$ CC and neutral current (NC) interactions in the material (liquid scintillator and iron of the support structure) of LVD.
\end{itemize}

We developed a full Montecarlo simulation that includes the generation of the neutrino interaction products, the propagation of the muon in the Gran Sasso rock and the response of the LVD detector. The details of the simulation were described in \cite{LVDmonitor}; however, with respect to that paper, some modifications were done with up--to--date informations. In particular we now use the CNGS flux calculated in 2005 by the Fluka group \cite{FlukaWeb} and the neutrino cross section NUX-FLUKA \cite{FlukaNux}. There are also some modifications in the detector: there are actually 7 active levels of scintillation counters instead of the 8 previously considered, and the energy threshold for the definition of a CNGS event is now $100$ MeV instead of $200$ MeV.

The resulting number of expected events, at the nominal intensity $4.5~10^{19}$ p.o.t./y is $32160$/y, equivalent to $7.147 ~10^{-16}$ events per p.o.t. (considering $200$ effective days per year, it corresponds to $\sim 160$ CNGS events per day): $78 \%$ are muons from the rock, $17 \%$ are CC interactions in the detector and $5 \%$ are NC.

During Run1 the total number of p.o.t. was $7.59~10^{17}$, thus $542$ events are expected in LVD.

\section{CNGS detected events}               %% As many sections as you want
\begin{figure}[h!]
\resizebox{0.5\textwidth}{!}{%
  \includegraphics{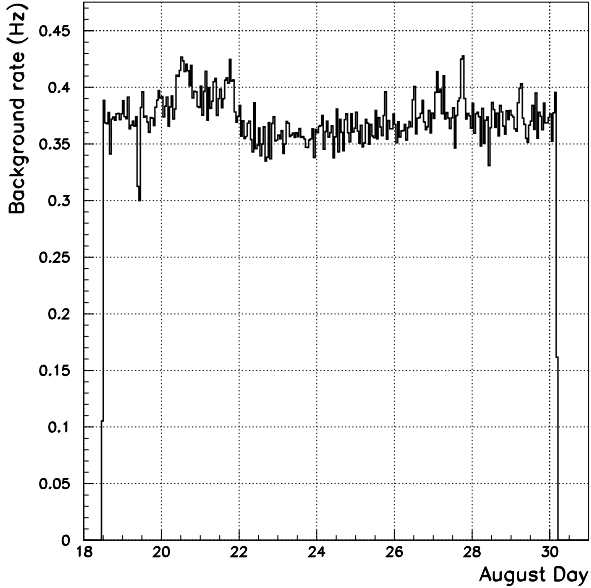}
}
%\mbox{\epsfig{file=bkg-rate.eps,height=10cm,width=10cm}}   %% EPS Figures
\caption{Background rate of events during Run1. These events present at least one scintillation counter with an energy release larger than $100$ MeV.}
\label{fi:bkgrate}

\end{figure}

The LVD events are filtered using a very loose selection cut: we require to have at least one scintillation counter with an energy release larger that $100$ MeV. The resulting rate is quite stable, with an average value of $0.37$ Hz, see figure \ref{fi:bkgrate}.

\nopagebreak

Among this sample the first selection criteria is based on the coincidence of the LVD event time with the beam spill time written in the DB. Two main corrections have been done: the neutrino time of flight from Cern to the LNGS ($2.440$ ms) and the propagation of the GPS time signal from the outside laboratories to slave clocks in the underground hall ($42116$, $42064$ and $42041$ ns respectively for tower 1, 2 and 3), measured in July, 2006 together with the other experiments at LNGS \cite{Autiero-clock}.

During Run1 there were some additional time shifts between the LNGS time and the UTC time written in the DB: from the beginning until $18^{th}$ 16:00 it was $+100 ~\mu$s, then $+10 ~\mu$s until $22^{nd}$ 9:00 and $-2$ ms until $23^{rd}$ 5:30. After the MD there was no additional bias.

\begin{figure}[h!]
\resizebox{0.5\textwidth}{!}{%
  \includegraphics{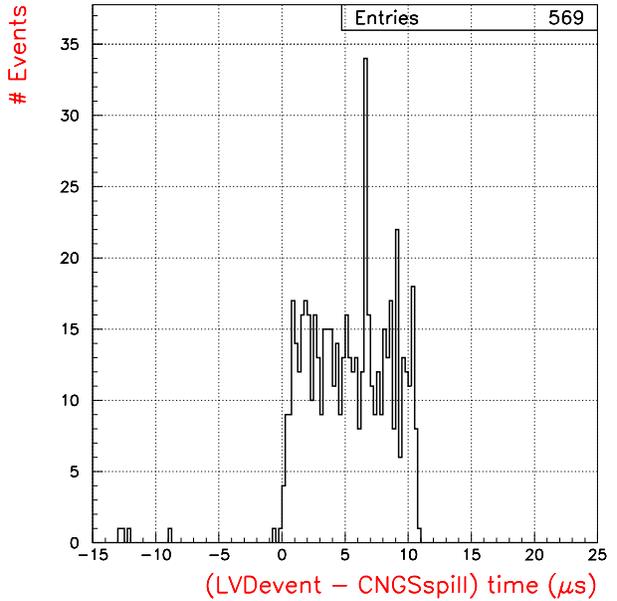}
}
%\mbox{\epsfig{file=time-in-spill.eps,height=10cm,width=10cm}}   %% EPS Figures
\caption{Distribution of the detection time of the 569 CNGS events, with respect to the initial time of the beam spill.}
\label{fi:gate}
\end{figure}

After applying all these corrections, we search for the CNGS events in the interval $[-15, +25]~ \mu$s around the start time of the beam spill. In this way 569 events are selected; their distribution in the time window is shown in figure \ref{fi:gate}. 

The four events with time difference between $-13$ and $-8~ \mu$s occured on 19$^{th}$ August when there was a failure in the LNGS master clock. The second master was switched on and a time shift of about $10 ~ \mu$s was present. Thus those events are considered in the analysis.

\nopagebreak

\begin{figure*}[t]
  \begin{minipage}{.40\columnwidth}
      \includegraphics[height=22pc]{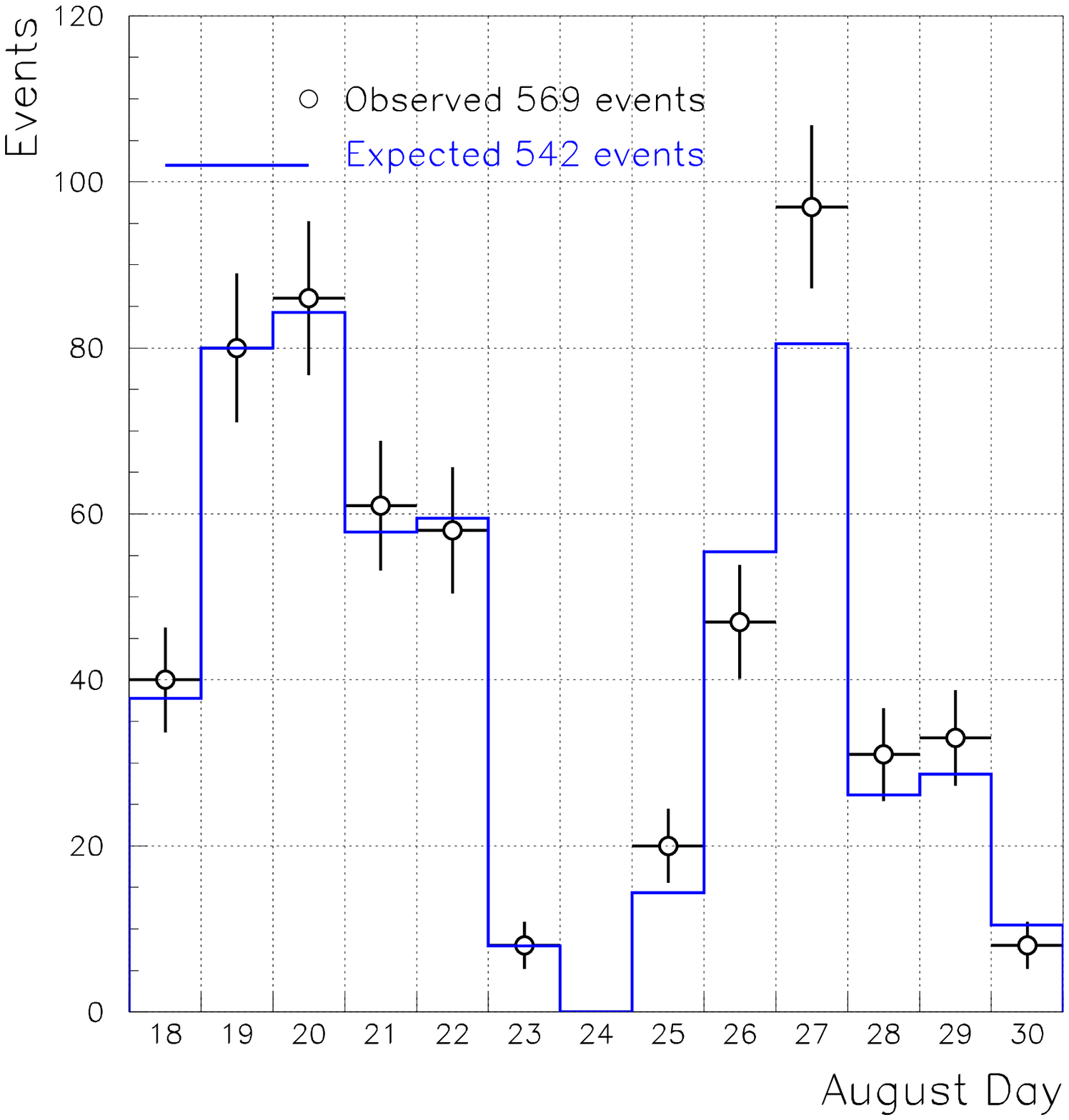}
%    \vspace{-1.cm}
  \end{minipage}
  \hspace{12.5pc} %%%%% space between two figures
  \begin{minipage}{.40\columnwidth}
      \includegraphics[height=22pc]{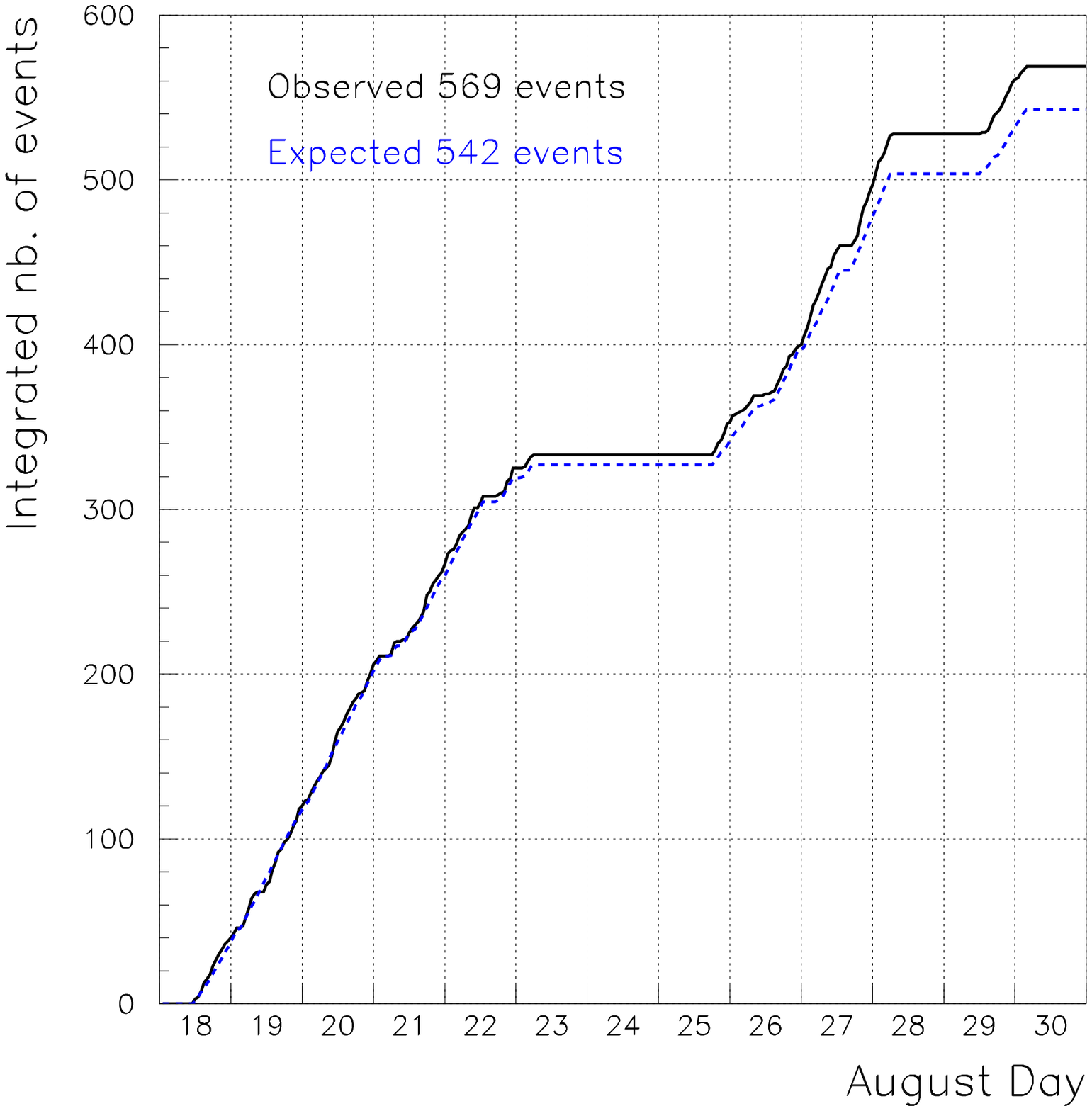}
  \end{minipage}
    \caption{Comparison between the expected and detected number of events during Run1. In the {\it left} panel there is the number of events per day: observed (black circles) and expected (blue line). In the {\it right} one the integrated number of events: observed (black solid line) and expected (blue dashed line).}
    \label{fi:compdiff}
\end{figure*}

In figure \ref{fi:compdiff} (left) we show the comparison between the expected and detected event rate per each day of data acquisition; in figure \ref{fi:compdiff} (right) the comparison of the integral number of events, hour by hour, is shown. Given the presently limited statistics, the agreement is rather good.

\begin{figure*}[t]
  \begin{minipage}{.70\columnwidth}
      \includegraphics[height=25pc]{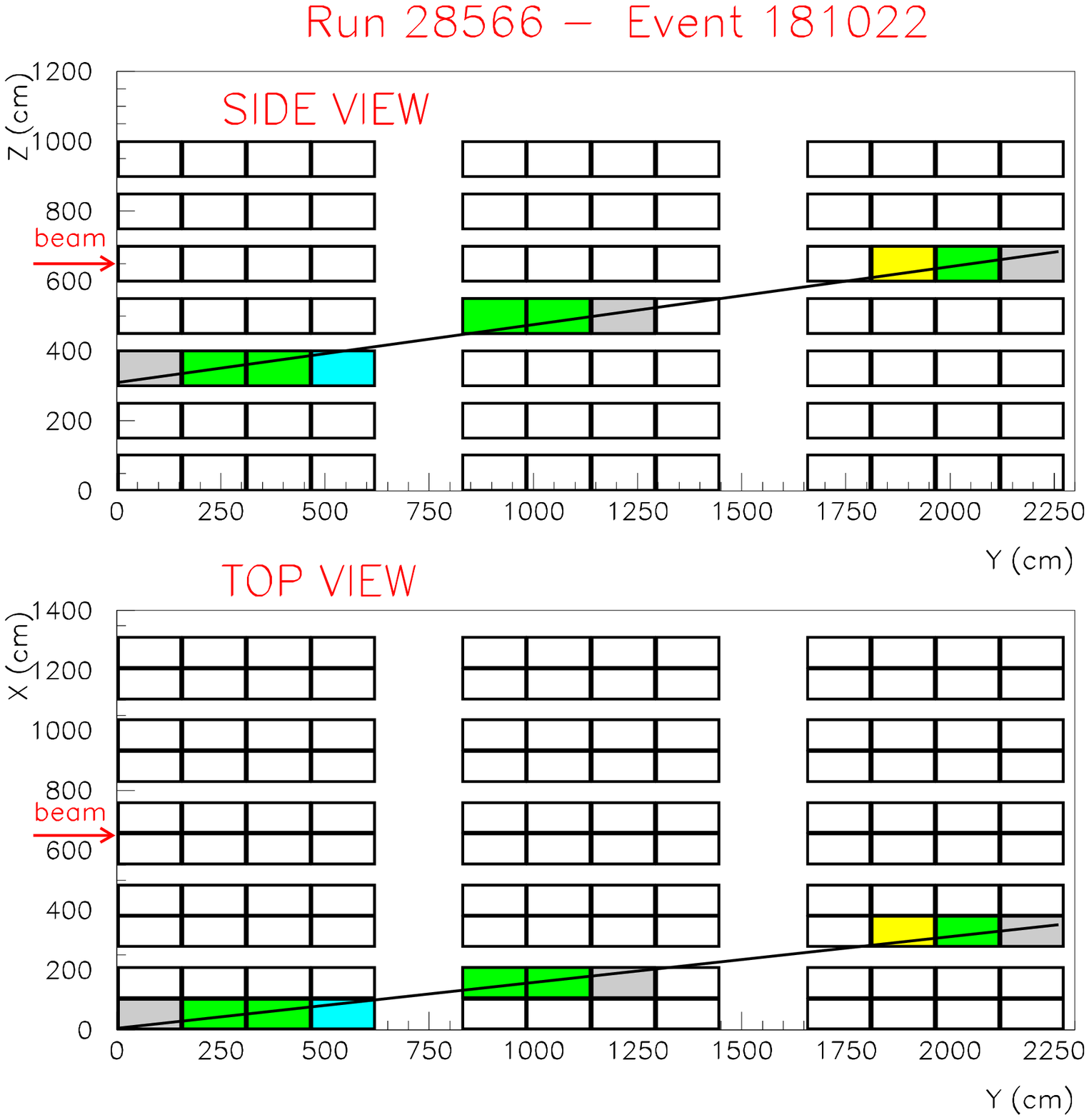}
%    \vspace{-1.cm}
  \end{minipage}
  \hspace{13.5pc} %%%%% space between two figures
  \begin{minipage}{.15\columnwidth}
      \includegraphics[height=10pc]{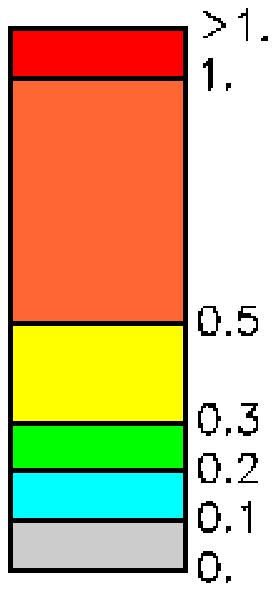}
  \end{minipage}
    \caption{Display of a CNGS events: typical charged current interaction in the rock upstream LVD, producing a muon that goes through the detector. The colours represent the amount of energy released in the scintillation counters, summed along each projected view; the legenda is expressed in GeV. The black straight line is the result of a linear fit to the hit counters.}
    \label{fi:dispmu}
\end{figure*}

\begin{figure}[t]
%  \begin{minipage}{.70\columnwidth}
\resizebox{0.5\textwidth}{!}{%
  \includegraphics{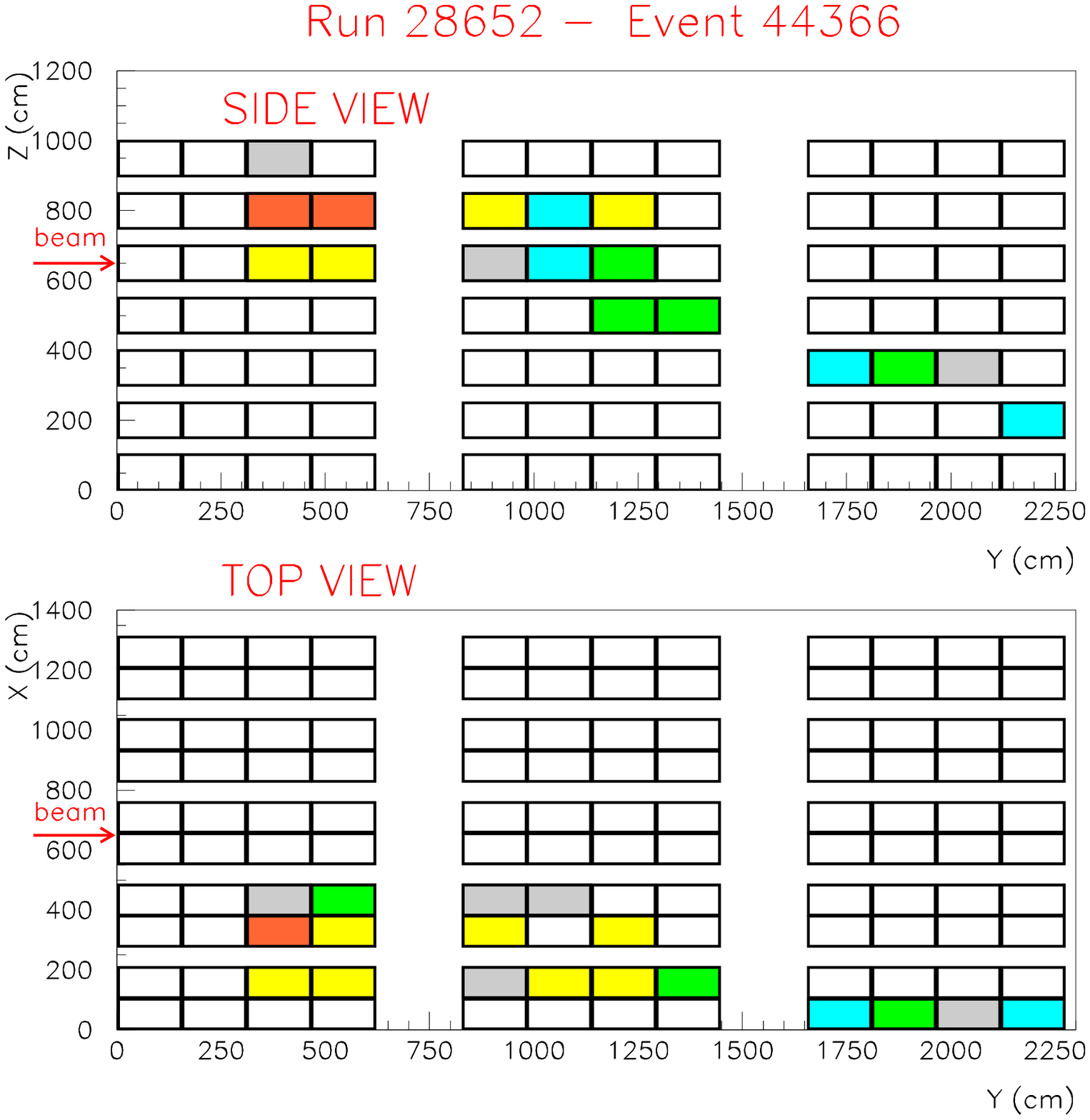}
}
%%      \includegraphics[height=20pc]{./display_28652_44366_2nd.eps}
%    \vspace{-1.cm}
%  \end{minipage}
%  \hspace{3.5pc} %%%%% space between two figures
%  \begin{minipage}{.15\columnwidth}
%      \includegraphics[height=10pc]{./EnergyScale.eps}
%  \end{minipage}
    \caption{Display of a CNGS events: neutrino interaction inside the LVD detector. The colours represent the amount of energy released in the scintillation counters, summed along each projected view; the legenda is the same as in figure \ref{fi:dispmu}.}
    \label{fi:dispnucc}
\end{figure}

Two examples of typical CNGS events in LVD are shown in figure \ref{fi:dispmu} (muon from the rock) and \ref{fi:dispnucc} (neutrino interaction inside the detector).

\subsection{Comparison with the MC simulation}
\begin{figure}[h!]
\begin{center}
\resizebox{0.5\textwidth}{!}{%
  \includegraphics{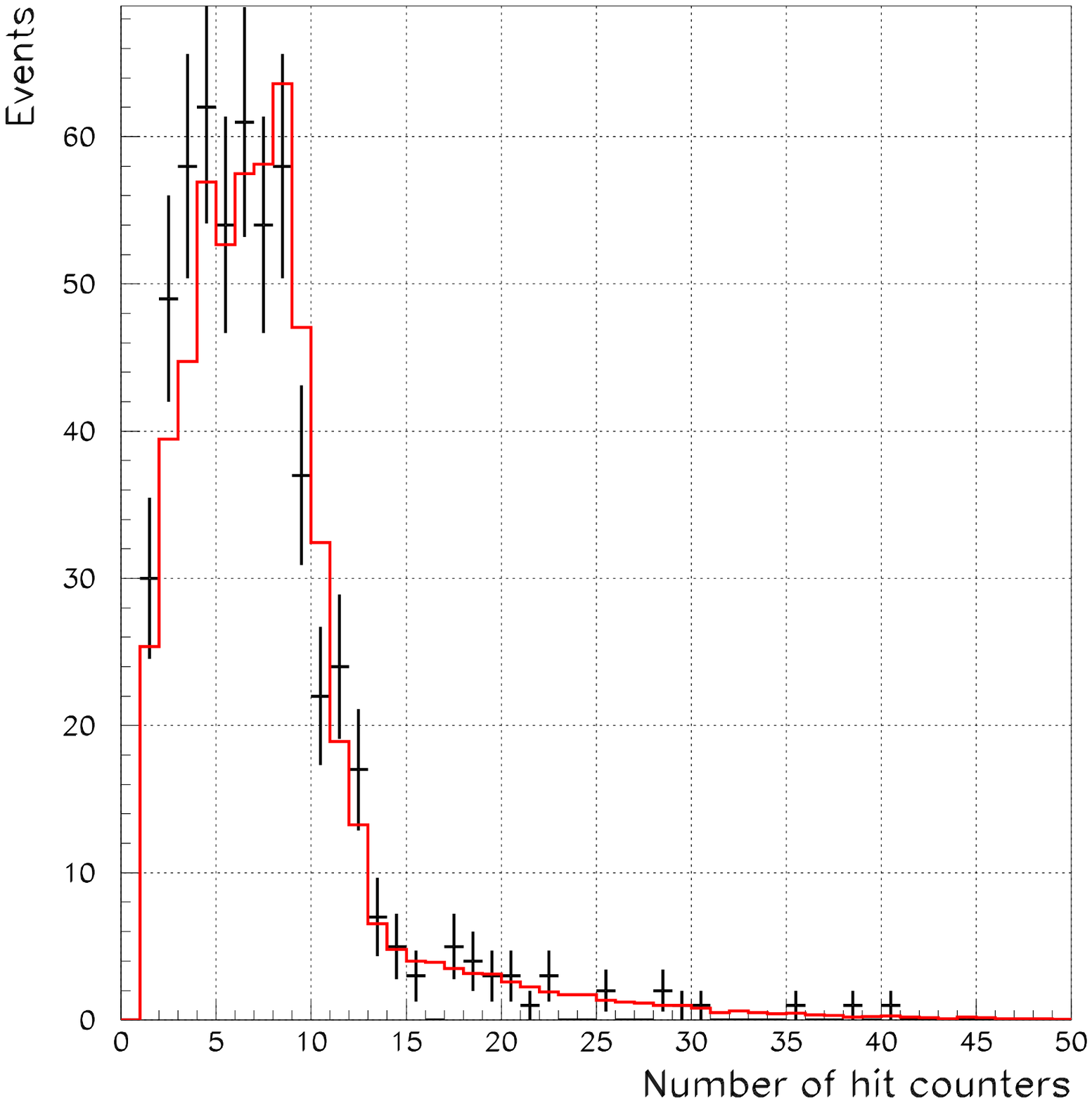}
}
\caption{Distribution of the number of scintillation counters hit per each CNGS event: comparison between the data (black crosses) and MC simulation (red line).}
\label{fi:ncounters}
\end{center}
\end{figure}

In the following we show the comparison between the CNGS detected events and the results of our MC simulation, normalized to the same number of events.
In figure \ref{fi:ncounters} we show the distribution of the number of scintillation counters hit per each event, with an energy threshold of $10$ MeV, normalized to the total sample of $569$ detected events; given the available statistics the agreement is good.
%while in figure \ref{fi:totene} the total amount of energy detected by the scintillation counters is displayed. 
%These two histograms are normalized to the total sample of $569$ detected events; given the available statistics the agreement is quite good.

In order to select the events generated in the rock, producing a penetrating muon inside the detector, we perform the muon track reconstruction with a linear fit to the centers of the hit scintillation counters. Requiring at least 3 hit counters and a good $\chi^2$ (probability larger than $1\%$ in both the {\it TOP} and {\it SIDE} projections) we select $319$ events. From the MC simulation we estimate that, using this selection cut, the efficiency to detect ``muons from the rock'' is about $80\%$ and the contamination of ``internal events'' is low (less than $5 \%$).

\begin{figure}[h!]
\begin{center}
\resizebox{0.5\textwidth}{!}{%
  \includegraphics{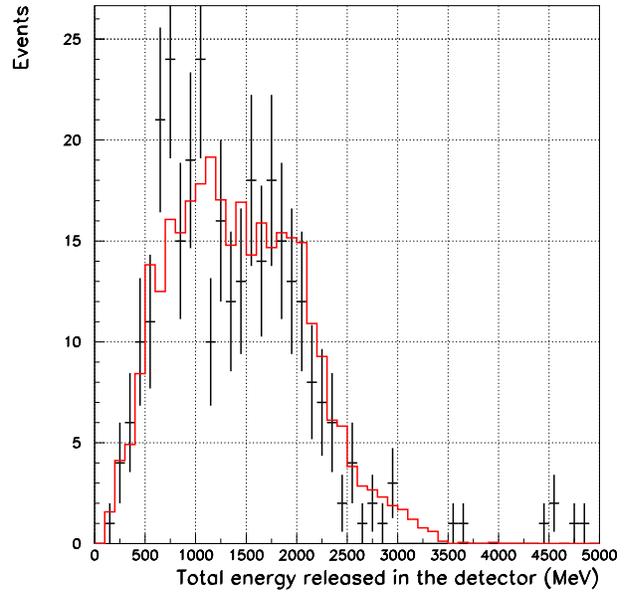}
}
\caption{Distribution of the total energy detected in the apparatus per each CNGS event where a muon is reconstructed: comparison between the data (black crosses) and MC simulation (red line).}
\label{fi:totene}
\end{center}
\end{figure}

In figure \ref{fi:totene} the total amount of energy detected by the scintillation counters in ``muonic'' events is displayed.

\begin{figure*}[h!]
  \begin{minipage}{.50\columnwidth}
      \includegraphics[height=10.cm]{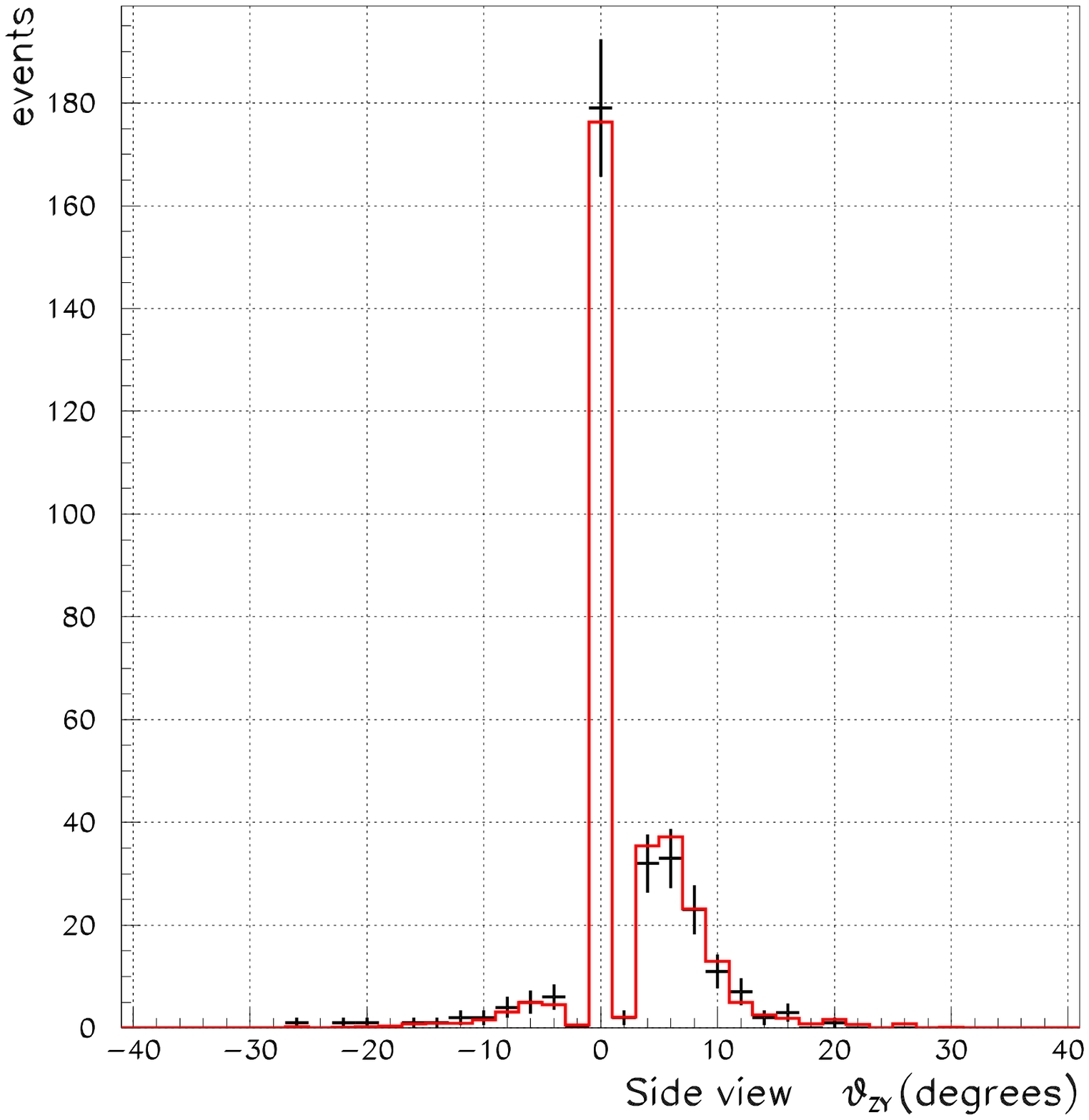}
%    \vspace{-1.cm}
  \end{minipage}
  \hspace{10.5pc} %%%%% space between two figures
  \begin{minipage}{.25\columnwidth}
      \includegraphics[height=4.5cm]{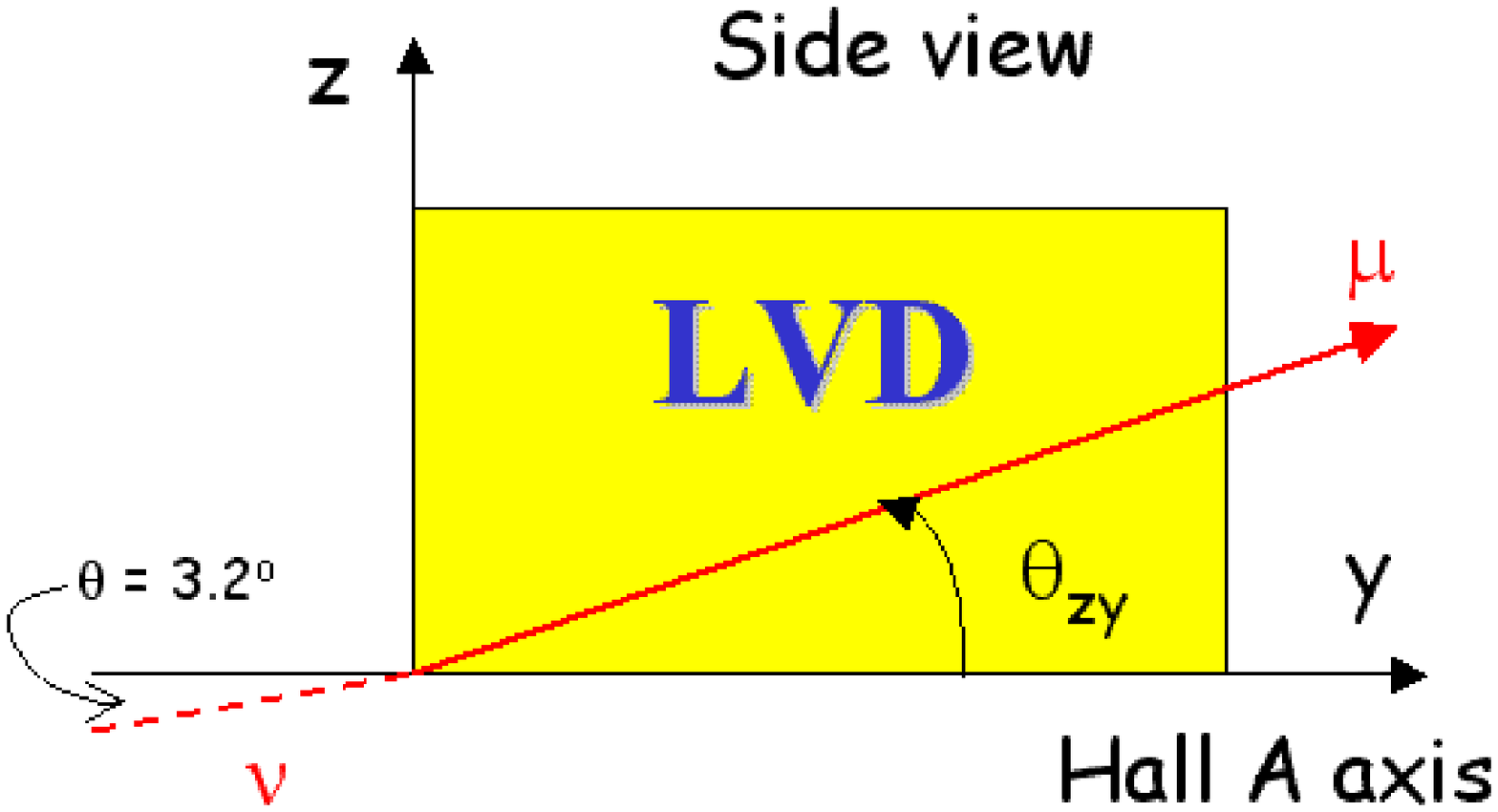}
  \end{minipage}
\caption{Distribution of the ``side'' projection of the angle between the reconstructed muon track and the main axis of the LNGS hall: comparison between the data (black crosses) and MC simulation (red line). In the right picture there is a description of the considered angle $\theta_{zy}$.}
\label{fi:murecside}
\end{figure*}

\begin{figure*}[h!]
  \begin{minipage}{.50\columnwidth}
      \includegraphics[height=10.cm]{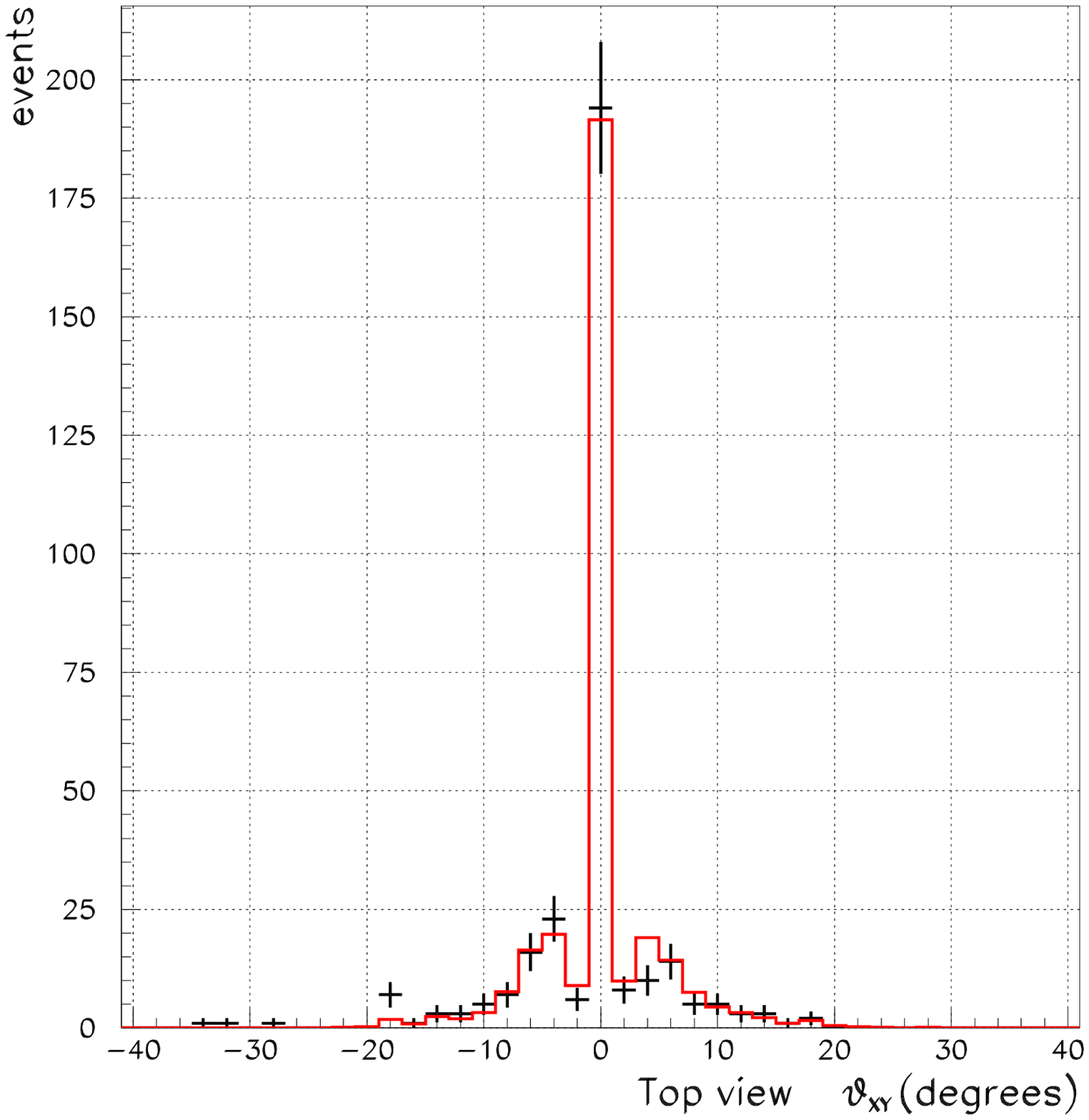}
%    \vspace{-1.cm}
  \end{minipage}
  \hspace{10.5pc} %%%%% space between two figures
  \begin{minipage}{.25\columnwidth}
      \includegraphics[height=4.5cm]{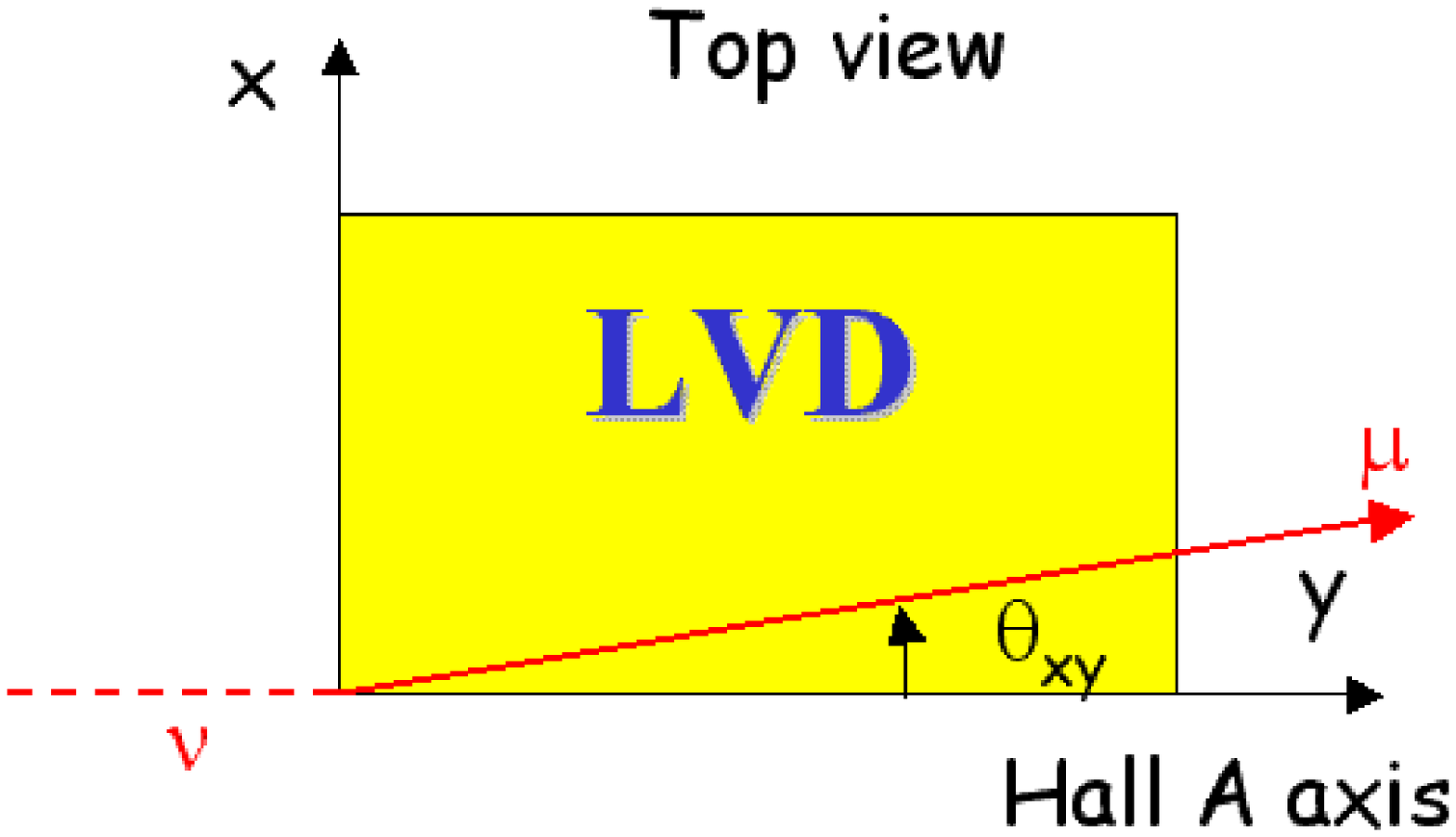}
  \end{minipage}
\caption{Distribution of the ``top'' projection of the angle between the reconstructed muon track and the main axis of the LNGS hall: comparison between the data (black crosses) and MC simulation (red line). In the right picture there is a description of the considered angle $\theta_{xy}$.}
    \label{fi:murectop}
\end{figure*}

For this selected sample of events we can reconstruct the muon direction and compare it with the expectation from the reconstruction of MC events.
The results are shown for the angle between the muon and the main axis of the hall A: its projections in the {\it SIDE} and {\it TOP} view of the detector are in figure \ref{fi:murecside} and \ref{fi:murectop} respectively. 

The events are almost horizontal and the main part of them are reconstructed exactely at $0^{\circ}$ because of the discreteness of the scintillation counters (cross section $1 \times 1$ m$^2$). In the {\it TOP} view the beam is parallel to the hall A axis, while in the {\it SIDE} view the beam ``comes out'' from the floor with an angle of $3.2^{\circ}$, as seen in figure \ref{fi:murecside}.
The agreement of the data and the MC simulation is very good in both the projections.

\subsection{Background}
The background is estimated considering the rate of events shown in figure \ref{fi:bkgrate} among which the CNGS events are searched for, with an average value of $0.37$ Hz.
We remind that the time window where we search for the events around the beam spill time is $40~\mu$s wide, and the number of useful spills in the DB is 51581. Thus the number of events due to the background, during Run1, is $$N_{bkg} = 0.37~ Hz \times 40~ \mu s \times 51581 = 0.764$$ practically negligible.

\section{Conclusions}
We presented the results of the first events detected by the LVD detector in coincidence with the CNGS beam. The first run of the CNGS beam was started in August 2006, with an overall number of $7.6~10^{17}$ protons delivered to the target. The LVD detector was fully operative during the whole run, with an average active mass of $950$ t.

LVD can detect the CNGS neutrinos through the observation of penetrating muons originated by $\nu_{\mu}$ CC interactions in the rock upstream the LNGS and through internal CC and NC neutrino interactions. 

The expected number of events, as predicted by our Montecarlo simulation, is $542$.
We searched for the CNGS events by looking at the time coincidence with the beam spill time; the number of detected events is $569$.

There is a good agreement, between the detected events and the MC simulation, in the distribution of the number of hit counters, the total energy released in the apparatus and the direction of the reconstructed muons.

We estimate that the number of events due to the background is lower than one in the whole Run1 time.

Thus, this first run of the CNGS beam confirmed that, as it was proposed in \cite{LVDmonitor}, LVD can act as a very useful CNGS beam monitor. 

Also the Opera collaboration reported about their measurements \cite{opera}: they detect 319 events against a prediction of 300, obtaining results very similar to ours. 
A detailed discussion about the comparison of expected and detect events in LVD is postponed to the next CNGS runs (scheduled in fall 2007), when a large number of events will be available and the beam characteristics will be better under control.

\section{Acknowledgements}                   
We thank G. Battistoni, A. Ferrari and P. Sala for providing us the tables of the neutrino cross-sections and A. Guglielmi for the details of the CNGS MC simulation.
We also would like to thank D. Autiero for his help in accessing the information in the CERN database and for many discussions about the details of the beam characteristics and status during Run1.

%
% BibTeX users please use
% \bibliographystyle{}
% \bibliography{}

\end{document}